\documentclass[12 pt]{article} 
\usepackage{hyperref}
\usepackage[utf8]{inputenc} 

\usepackage[a4paper, total={6in, 8in}, margin=1in]{geometry}
\usepackage{float}
\usepackage{listings}
\usepackage{url}
\usepackage[table,xcdraw]{xcolor}
\usepackage{graphicx} 
\usepackage{booktabs} 
\usepackage{array} 
\usepackage{amsmath,mathtools}
\usepackage{enumitem}
\usepackage[format=plain, labelfont={color=darkgray, bf}, textfont={color=darkgray}]{caption}
\usepackage{verbatim} 
\usepackage{subfig} 
\usepackage{fancyhdr} 
\usepackage{titling}
\usepackage[nottoc,notlof,notlot]{tocbibind} 
\usepackage{gensymb}
\usepackage{verbatim}
\usepackage{soul}
\let\oldtextbf\textbf
\renewcommand{\textbf}[1]{\oldtextbf{\boldmath #1}}
\usepackage{svg}
\usepackage{tcolorbox}

\title{constraints of anisotropy on warm power-law inflation in light of Planck results }
\author{Zahra Ghadiri${^1}$ , Ali Aghamohammadi${^2}$  and Abdollah Refaei  
\\Sanandaj Branch, Islamic Azad University, Sanandaj, Iran
\\$1.$ zghadiry1@gmail.com
\\$2.$ a.aghamohamadi@iausdj.ac.ir }

\begin{document}

\maketitle

\textbf{Abstract}\\

\quad
     In this article, we examine the effects of anisotropy on a model of canonical warm inflation with the power-law potential.  The results of this model are compared to Planck satellite observational data. Using a conventional local scalar field in the Bianchi type I metric, the slow-roll conditions and the suitable regions, where the free parameters of the model show a good agreement with Planck results, are investigated in detail. Following the usual calculations for warm inflationary approaches, the early universe in two different dissipative regimes, namely the weak dissipative and the strong dissipative ones, is investigated. In this regard, the slow-roll parameters and their observational indices in both regimes are obtained and finally, it is shown that this model is in good agreement with observations.

Keywords:Warm power-law inflation, Weak and strong dissipative regimes, Planck data, Anisotropy.


\section{Introduction }

\quad The cosmic inflationary scenario provides a desirable mechanism for describing the large-scale structures of the Universe\cite{1}$-$\cite{7}.   We have studied different models of inflation in the sources\cite{8}\cite{9}. And among all those models, we found that there is a general division called cold inflation and warm inflation. More details are in Ref [\cite{10}$-$\cite{21}]. The main condition for generating warm inflation is that (T $>$ H), where T refers to the flow temperature which indicates thermal fluctuations and H means  the Hubble parameter which includes quantum fluctuations([\cite{22}$-$\cite{26},\cite{12}$-$\cite{15}]. We know that quantum and thermal fluctuations play an important role in the formation of initial cosmic density fluctuations and large-scale structures\cite{13}$-$\cite{18},\cite{27}. In 1983, the theory of chaotic inflation was introduced following Linde's studies in which the power-law potentials were introduced, and today we see that the power-law potentials due to simplicity and compliance with observations and solve the problem of graceful exit from the horizon are considered in new models\cite{28} Two structures are commonly used in the calculation of slow-roll parameters in inflation models; One is the structure in which the parameters are expressed in terms of the potential of the scalar field, and this is the structure that we have discussed in this article and Ref\cite{29}.And the other is the structure in which the parameters are expressed in terms of the Hubble parameter and is called the Hamilton Jacobi form. So far, many models in the framework of gravity theories,\cite{30} and observational results related to CMB and the formation of large-scale structures measured by advanced instruments have been able to provide us with useful information  \cite{31}$-$\cite{18}.Among them is the existence of statistical anisotropy in the CMB spectrum, which has raised doubts about the existence of anisotropies in the inflation period. What has been said so far is the theoretical and observational reasons for the importance of addressing the issue of inflation, and especially the issue of anisotropy in a period of evolution universe, but it is not yet clear what kind of gravity existed in the period of inflation. Therefore, in this study, to help advance the studies, we decided to investigate the existence of anisotropy in a model of cosmic warm inflation in the Bianchi type I framework. It should be noted that the Bianchi model is an extension of the standard FLRW flat model and is the simplest model in the world that depicts an anisotropic but homogeneous flat space\cite{39}$-$\cite{40} Compared to the world FLRW, which has the same scale factor for its spatial directions. In the Bianchi world, the scale factor can change in different independent directions. For the sake of  the importance of the subject and the fact that the study of anisotropic inflation world has better advantages than the isotropic world, the present study has addressed this issue. Accordingly, we compare our model calculations with the latest observational data from the observations of the Wilkinson and Planck microwave anisotropy probe satellite, including the amplitude of scalar spectrum perturbations and the scalar and tensor spectral indices. In the second part of this article, we have stated the theoretical framework of anisotropic warm inflation. In the third section, we examine the effects of anisotropy in the weak dissipative regime and identify the best areas for the appropriate parameters. In the fourth section, we examine the anisotropic effects on the strong dissipative regime, and similarly to the third section, we obtain the areas where the parameters can best match with the observations. Finally, in the fifth section, we have summarized the results of our final reviews and comments.

\section{	General Framework }

The BI metric is expressed as follows\cite{39}

\begin{equation}\label{1} 
{ds}^2={dt}^2-A^2\left(t\right){dx}^2-B^2\left(t\right){(dy}^2+{dz}^2),                    
\end{equation} 

The coefficients A, B, C are a function of  t.  A  is the scale factor in the direction of the x-axis and B is the scale factor in the direction of the y and z axes. More details on the Bianchi metric are given in Ref\cite{39}$-$\cite{40}. Because Planck data show CMB anisotropy that could be the reason why the world is not completely isotropic therefore, we considered the anisotropy as small deviation of isotropy in the Bianchi model, that is an extension of the standard FLRW flat model and is the simplest model in the world that depicts an anisotropic but homogeneous flat space.\\
The following are brings some field equations related to the BI metric:\cite{41}$-$\cite{42}

\begin{equation} \label{2}
H_{tot}=\frac{1}{3}\left(H_1+2H_2\right),\ \ \ \ \ \ \ \ \ 
\end{equation}

\begin{equation} \label{3}
H_1=\frac{\dot{A}\left(t\right)}{A\left(t\right)},\ \ \ \ \ \ \ \ \ H_2=H_3=\frac{\dot{B}(t)}{B(t)},\ \ A=B^{\lambda }\ \ \ \ \ ,\ H_1=\lambda H_2,\
\end{equation} 
\begin{equation} \label{4} 
3{H_{tot}}^2-{\sigma }^2=\frac{1}{M^2_P}({\rho }_{\mathit{\Phi}}+{\rho }_r)\ ,                                                                 
\end{equation}

\begin{equation} \label{5}
\mathrm{3}{H_{tot}}^{\mathrm{2}}\mathrm{+2}\dot{H_{tot}}\mathrm{+}{\sigma }^{\mathrm{2}}\mathrm{=-}\frac{\mathrm{1}}{M^{\mathrm{2}}_P}P\left(\mathrm{\ }\phi \mathrm{\ }\right)
 \ \ \ \ \ 
\end{equation}

The equations for the Hubble parameter and the shear parameter are expressed as follows\cite{39}$-$\cite{40}

\begin{equation} \label{6}
H\mathrm{=}\frac{\mathrm{1}}{\mathrm{3}}\left(\lambda \mathrm{+2}\right)H_{\mathrm{2}}
\end{equation}

\begin{equation} \label{7}
{\sigma }^{\mathrm{2}}\mathrm{=}\frac{{\left(\lambda \mathrm{-}\mathrm{1}\right)}^{\mathrm{2}}H^{\mathrm{2}}_{\mathrm{2}}}{\mathrm{3}}
\end{equation}

It should be noted that in the equations given above,$ \lambda$ is a constant value and shows the degree of deviation from isotropy along the x-axis. If the value of $\lambda$$=$1, we will have the same FLRW space-time flat with signature$ -$2.
As we know, the energy density and pressure of the scalar field and the density of the radiation field are defined as follows:\cite{29}

\begin{equation} \label{8}
{\rho }_{\phi }\mathrm{=}\frac{\mathrm{1}}{\mathrm{2}}{\dot{\phi }}^{\mathrm{2}}\mathrm{+}V\left(\phi \right)
\end{equation}

\begin{equation} \label{9}
P_{\phi }\mathrm{=}\frac{\mathrm{1}}{\mathrm{2}}{\dot{\phi }}^{\mathrm{2}}\mathrm{-}V\left(\phi \right)
\end{equation}

\begin{equation} \label{10}
{\rho }_r\mathrm{=}\frac{{\pi }^{\mathrm{2}}}{\mathrm{30}}g_{\mathrm{*}}T^{\mathrm{4}}\mathrm{=\ }\mathrm{c}\mathrm{\ }T^{\mathrm{4}}
\end{equation}

Here, T is fluid temperature; $c$ is the Stefan Boltzmann constant, and the value of c $=$ 70. Meanwhile, the equation of the anisotropic warm inflation field is expressed as follows: \cite{29},  \cite{12}$-$\cite{15}

\begin{equation} \label{11}
\ddot{\phi }\mathrm{+}H_{\mathrm{2}}\left(\lambda \mathrm{+2}\right)\left(\mathrm{1+}R\right)\dot{\phi }\mathrm{+}V_{,\phi }\mathrm{=0,
}
\end{equation}

Where $R\mathrm{\equiv}\frac{\mathrm{\Gamma }}{\left(\lambda \mathrm{+2}\right)H_{\mathrm{2}}}$ is dissipation function and   
$\Gamma{ (T,\phi)}\mathrm{=}{a}{T^{n}}{\phi^{1-n}}$ is called the dissipation coefficient. The dissipative coefficient can be a function of time, a function of the scalar field, or a function of both \cite{27}$-$\cite{28}
In this case, the conservation equations of ${\rho }_{\phi }$  ,$ P_{\phi }$ are described by the following equations

\begin{equation} \label{12}
\dot{{\rho }_{\phi }}\mathrm{+}\left(\lambda \mathrm{+}\boldsymbol{\mathrm{2}}\right)H_{\boldsymbol{\mathrm{2}}}\left({\rho }_{\phi }\mathrm{+\ }P_{\phi }\right)\mathrm{=-}\mathrm{\Gamma }{\dot{\phi }}^{\boldsymbol{\mathrm{2}}}
\end{equation}

\begin{equation} \label{13} 
\dot{{\rho }_r}\mathrm{+}\frac{\mathrm{4}}{\mathrm{3}}\left(\lambda \mathrm{+2}\right)H_{\mathrm{2}}{\rho }_r\mathrm{=}\mathrm{\Gamma }{\dot{\phi }}^{\mathrm{2}},                                               
\end{equation}

Applying the conditions of slow- Roll the Eqs .($\eqref{11}$ and $\eqref{13}$)are expressed approximately as

\begin{equation} \label{14}
{\rho }_r\mathrm{=}\alpha \mathrm{\ }T^{\mathrm{4}}\mathrm{\simeq }\frac{\mathrm{3}\mathrm{\Gamma }{\dot{\phi }}^{\mathrm{2}}}{\mathrm{4}\left(\lambda \mathrm{+2}\right)H_{\mathrm{2}}}
\end{equation} 

\begin{equation} \label{15}
H_{\mathrm{2}}\left(\lambda \mathrm{+2}\right)\left(\mathrm{1+}R\right)\dot{\phi }\mathrm{+}V_{,\phi }\mathrm{=0,}
\end{equation}

The necessary condition for the occurrence of warm inflation is that, its slow-roll  parameters are defined as follows.

\begin{equation} \label{16} 
\epsilon \mathrm{=}\frac{M^{\mathrm{2}}_P}{\mathrm{2}}{\left(\frac{V_{,\phi }}{V}\right)}^{\mathrm{2}}\mathrm{\ \ \ \ ,\ \ \ \ \ \ \ \ }\eta \mathrm{=\ }M^{\mathrm{2}}_P\mathrm{\ \ }\left(\frac{V_{,\phi \phi }}{V}\right)\ \ \ \ \ \ \ 
\end{equation}

\begin{equation} \label{17} 
\epsilon \ll 1+R\ \ ,\ \ \ \ \ \ \ \ \eta \ \ll 1+R\ \ 
\end{equation}

Another important parameter that should be calculated in warm inflation is the number of e-folds which is an effective role in solving the horizon problem and is expressed as follows:

\begin{equation} \label{18} 
N\mathrm{=-}\int^{\mathit{\Phi}}_{{\mathit{\Phi}}_{end}}{\frac{\frac{\mathrm{1}}{\mathrm{3}}\left(\lambda \mathrm{+2}\right)H_{\mathrm{2}}}{\dot{\mathit{\Phi}}}\mathrm{\ }d\mathit{\Phi}}
\end{equation}

To estimate the best free parameters in any inflation model, it is required to obtain the amplitude of scalar and tensor perturbations, tensor-to-scalar spectrum ratio, the scalar spectral index and running parameter, based on  the definition of anisotropic warm inflation, hence,  its equations are defined as follows:

\begin{equation} \label{19} 
{\mathcal{P}}_s=\frac{25\ H^2}{4{\dot{\ \mathit{\Phi}}}^2}{\delta \mathit{\Phi}}^2=\frac{\ {25\left(\lambda +2\right)}^2{H_2}^2}{36{\dot{\ \mathit{\Phi}}}^2}{\delta \mathit{\Phi}}^2
\end{equation} 

\begin{equation} \label{20} 
{\mathcal{P}}_t=\frac{2\ {\left(\lambda +2\right)}^2{H_2}^2}{\ 9\ {\pi }^2M^2_P},\  
\end{equation}

\begin{equation} \label{21} 
r\equiv \frac{{\mathcal{P}}_t}{{\mathcal{P}}_s},\ 
 \ \ \end{equation} 

\begin{equation} \label{22} 
n_s-1\equiv \frac{d{\mathrm{ln} {\mathcal{P}}_s\ }}{d{\mathrm{ln} k\ }}\ \ \ \
\end{equation} 

\begin{equation} \label{23} 
{\alpha }_s\mathrm{=}\frac{dn_s}{dlnk}
\end{equation}

From Ref \cite{43} we obtained:

\begin{equation} \label{24} 
a\frac{\left(\lambda \mathrm{+2}\right)}{\mathrm{3}}H_{\mathrm{2}}\mathrm{=}c_sk \mathrm{,\ \ \ \ \ \ \ \ \ \ \ \ }\frac{d}{d\mathrm{\ }lnk}\mathrm{\simeq }\mathrm{-}\frac{d}{dN}
\end{equation}

In the following, we will inspect the effects of anisotropy in a model of warm inflation with the power-law potential  in two strong and weak dissipative regimes, and observe that in the low dissipation regime the dominant parameter is the Hubble parameter, while in the strong  dissipation regime the  dissipation coefficient is the dominant parameter  . We are interested to find out the best fit for the free parameters in our proposed model.

\section{	Investigation of Effects Anisotropy in Weak Dissipative Regime}

In this case,  we try to test our model on a weak dissipative regime for R$ <$ 1 and potential  $V\left(\mathit{\Phi}\right)\mathrm{=}V_{0}{\Phi^{0}}$ . In slow-roll approximation, the Friedman equations and the temperature equation will be as follows:

\begin{equation} \label{25} 
{H_{\mathrm{2}}}^{\mathrm{2}}\mathrm{=}\frac{\mathrm{1}}{\left(\mathrm{2}\lambda \mathrm{+1}\right)M^{\mathrm{2}}_P}\mathrm{\ }V\left(\mathit{\Phi}\right)
\end{equation}

\begin{equation} \label{26} 
H_{\mathrm{2}}\left(\lambda \mathrm{+2}\right)\dot{\phi }\mathrm{+}V_{,\phi }\mathrm{=0,\ \ \ } 
\end{equation} 

\begin{equation} \label{27}
T\mathrm{=}\gamma \mathrm{\ \ }{\phi }^{\frac{\mathrm{-}\mathrm{1+}\frac{k}{\mathrm{2}}\mathrm{-}n}{\mathrm{4-}n}\mathrm{\ \ \ \ \ \ \ \ \ \ \ \ \ \ \ \ \ \ }}\mathrm{,\ \ \ \ \ \ }\gamma \mathrm{=}{\left(\frac{\mathrm{3}}{\mathrm{4}}\right)}^{\frac{\mathrm{1}}{\mathrm{4-}n}}{\left(\frac{a\mathrm{\ }M^{\mathrm{2}}_Pk^{\mathrm{2}}\mathrm{\ }{\left(1\mathrm{+2}\lambda \right)}^{\frac{\mathrm{3}}{2}}\mathrm{\ }{V_0}^{\frac{\mathrm{1}}{2}}}{\alpha \mathrm{\ }{\left(\mathrm{2+}\lambda \right)}^{\mathrm{3}}}\mathrm{\ }\right)}^{\frac{\mathrm{1}}{\mathrm{4-}n}}\
\end{equation}

The first parameter of the slow-roll is acquired  as:
\begin{equation} \label{28} 
\epsilon \mathrm{=}\frac{k^{\mathrm{2}}M^{\mathrm{2}}_p}{\mathrm{2}{\phi }^{\mathrm{2}}}\mathrm{\ ,\ \ } 
\end{equation}

As we know, inflation ends when, $\varepsilon$$=$1 and from that,  the measured field is at the end of inflation as follows

\begin{equation} \label{29} 
{\phi }_{end}\mathrm{=}\frac{kM_p}{\sqrt{\mathrm{2}}},                                 
\end{equation} 

The field function at the moment of crossing the horizon is also obtained  by using \cite{18}  as follows 
\begin{equation}\label{30}
\phi \mathrm{=}\frac{\sqrt{\mathrm{(-2}k\mathrm{\ }M^{\mathrm{4}}_p\mathrm{\ }N\mathrm{)(\ \ }1\mathrm{+2\ }\lambda )\mathrm{+(2+}\lambda \mathrm{)\ }{{\phi }_{end}}^{\mathrm{2}}}}{\sqrt{\frac{\mathrm{2}k\mathrm{+5}k\mathrm{\ }\lambda \mathrm{+2}k\mathrm{\ }{\lambda }^{\mathrm{2}}}{k(1\mathrm{+2}\lambda )}}}
\end{equation}

In the following, we obtain the perturbation parameters and compare them with the values obtained from the observations. The amplitude of the scalar spectrum perturbations and the scalar and tensor spectral, tensor to scalar ratio, and running, are  equations that are defined as :\cite{22}, \cite{29}

From $\eqref{19}$ , In weak dissipative regime with  
${\sigma}{\phi^{2}}\mathrm{\cong}\frac{\mathrm{1 }}{\mathrm{3}}{(\lambda \mathrm{+2})}H_{\mathrm{2}}{T}$
We have:

\begin{equation}\label{31}
\begin{aligned}
&\mathcal{P}_{\mathrm{s}}=\\
& \dfrac{25 \times 2^{\frac{46+\mathrm{k}-14 \mathrm{n}}{4(-4+\mathrm{n})}} \gamma(2+\lambda)^{2} \mathrm{M}_{\mathrm{p}}^{\frac{14+\mathrm{k}-6 \mathrm{n}}{8-2 \mathrm{n}}+\frac{\mathrm{k}}{2}-5}\left(\frac{\mathrm{k}\left(\mathrm{k}(2+\lambda)-4 \mathrm{M}_{\mathrm{p}}^{2} \mathrm{~N}(1+2 \lambda)\right)}{2+\lambda}\right)^{\frac{14+\mathrm{k}-6 \mathrm{n}}{2(8-2 \mathrm{n})}} \sqrt{\frac{2\left(\frac{-\mathrm{k}}{2}\right)_{\mathrm{V}_{0}}\left(\frac{\mathrm{k}\left(\mathrm{k}(2+\lambda)-4 \mathrm{M}_{\mathrm{p}}^{2} \mathrm{~N}(1+2 \lambda)\right)}{2+\lambda}\right)^{\frac{\mathrm{k}}{2}}}{(\mathrm{k}(1+2 \lambda))^{2}}}}{(k(1+2 \lambda))^2}
\end{aligned}
\end{equation}

\begin{equation}\label{32}
\begin{aligned}
&a (n,k,N)= \dfrac{\alpha ( \lambda +2)^3}{M_p^2 k^2 (1+2 \lambda)^{\frac{3}{2}} \sqrt{v_0}} \times \\
& {\mathrm{(}\frac{{\mathcal{P}}_sM^{\mathrm{4+}\frac{\mathrm{14+}k\mathrm{-}\mathrm{6}n}{\mathrm{8-2}n}+\frac{\mathrm{k-2}}{\mathrm{2}}}_p{\left(k\mathrm{+2}k\lambda \right)}^{\mathrm{2}}}{\mathrm{25}{\mathrm{(}\frac{\mathrm{3}}{\mathrm{4}}\mathrm{)}}^{\frac{\mathrm{1}}{\mathrm{4-}n}}{\mathrm{\ 2}}^{\frac{\mathrm{46+}k\mathrm{-}\mathrm{14}n}{\mathrm{4(-4+}n\mathrm{)}}}{\mathrm{(2+}\lambda \mathrm{)}}^{\mathrm{2}}{\mathrm{(}\frac{\sqrt{k\mathrm{(}k\left(\mathrm{2+}\lambda \right)\mathrm{-}\mathrm{4}M^{\mathrm{2}}_pN\mathrm{(1+2}\lambda \mathrm{))}}}{\sqrt{\mathrm{2+}\lambda }}\mathrm{)}}^{\frac{\mathrm{14+}k\mathrm{-}\mathrm{6}n}{\mathrm{8-2}n}}\sqrt{\frac{{\mathrm{2}}^{\frac{\mathrm{-}k}{\mathrm{2}}}\mathrm{\ }V_0{\mathrm{(\ }\frac{\sqrt{k\mathrm{(}k\left(\mathrm{2+}\lambda \right)\mathrm{-}\mathrm{4}M^{\mathrm{2}}_pN\mathrm{(1+2}\lambda \mathrm{)}}}{\sqrt{\mathrm{2+}}\lambda }\mathrm{\ )}}^K}{\mathrm{(1+2}\lambda \mathrm{)}}}}\mathrm{)}}^{\mathrm{4-}n}
\end{aligned}
\end{equation}

\begin{equation} \label{33} 
n_s\mathrm{(}n\mathrm{,\ }k,N\mathrm{)=1+}\frac{M^{\mathrm{2}}_p\left(\mathrm{-}\mathrm{14+}k\mathrm{\ }\left(\mathrm{-}\mathrm{5+}n\right)\mathrm{+6}n\right)\mathrm{(1+2}\lambda \mathrm{)}}{\mathrm{(-4+}n\mathrm{)}B\mathrm{(}k\left(\lambda \mathrm{+2}\right)\mathrm{-}\mathrm{4\ }M^{\mathrm{2}}_pN\left(\mathrm{1+2}\lambda \right)\mathrm{)}}, 
\end{equation} 

\begin{equation} \label{34} 
{\alpha }_{s\mathrm{\ }}\left(n\mathrm{,\ }k,N\right)\mathrm{=-}\frac{\mathrm{4}{\mathrm{\ }M}^{\mathrm{4}}_p\mathrm{\ }\left(\mathrm{-}\mathrm{14+}k\left(\mathrm{-}\mathrm{5+}n\right)\mathrm{+6}n\right){\left(\mathrm{1+2}\lambda \right)}^{\mathrm{2}}}{\left(\mathrm{-}\mathrm{4+}n\right){\left(k\left(\mathrm{2+}\lambda \right)\mathrm{-}\mathrm{4\ }{\mathrm{\ }M}^{\mathrm{2}}_pN\left(\mathrm{1+2}\lambda \right)\right)}^{\mathrm{2}}}
\end{equation}

\begin{equation} \label{35} 
{\mathcal{P}}_{\mathrm{t}}\mathrm{=}\frac{\mathrm{2\ \ }V_0\mathrm{\ }{\mathrm{(2+}\lambda \mathrm{)}}^{\mathrm{2}}{\mathrm{(}\frac{k^{\mathrm{2}}{\mathrm{\ }M}^{\mathrm{2}}_p}{\mathrm{2}}\mathrm{+}\frac{\mathrm{6\ }k\mathrm{\ }{\mathrm{\ }M}^{\mathrm{2}}_{p\mathrm{\ }}N\mathrm{(1+2}\lambda \mathrm{)}}{{\mathrm{(2+}\lambda \mathrm{)}}^{\mathrm{2}}}\mathrm{)}}^{\frac{k}{\mathrm{2}}}}{\mathrm{9\ }{\mathrm{\ }M}^{\mathrm{4}}_p\mathrm{\ }{\pi }^{\mathrm{2}}\mathrm{(1+2}\lambda \mathrm{)}} 
\end{equation}

$r\left(n,k,N\right)\mathrm{=} $

\begin{equation} \label{36} 
 \frac{{\mathrm{2}}^{\frac{\mathrm{62-7}k\mathrm{-}\mathrm{18}n\mathrm{+2}kn}{\mathrm{16-4}n}}k^{\mathrm{2}}V_0\left(\mathrm{1+2}\lambda \right){\mathrm{\ }M}^{\frac{\mathrm{14+}k\mathrm{-}\mathrm{6}n}{\mathrm{2}\left(n\mathrm{-}\mathrm{4}\right)}-\frac{\mathrm{k-2}}{\mathrm{2}}+k}_p{\left(\frac{\sqrt{k\mathrm{\ }\left(k\left(\mathrm{2+}\lambda \right)\mathrm{-}\mathrm{4}{\mathrm{\ }M}^{\mathrm{2}}_pN\left(\mathrm{1+2}\lambda \right)\right)}}{\sqrt{\mathrm{2+}\lambda }}\right)}^{\frac{\mathrm{14+}k\mathrm{-}\mathrm{6}n}{\mathrm{2}\left(n\mathrm{-}\mathrm{4}\right)}}{\left(k\mathrm{\ \ }\left(k\mathrm{+}\frac{\mathrm{12}\left(N\mathrm{+2}N\lambda \right)}{{\left(\mathrm{2+}\lambda \right)}^{\mathrm{2}}}\right)\right)}^{\frac{k}{\mathrm{2}}}}{\mathrm{225\ }{\pi }^{\mathrm{2}}\gamma \sqrt{\frac{{\mathrm{2}}^{\frac{\mathrm{-}k}{\mathrm{2}}}V_0{\left(\frac{\sqrt{k\mathrm{(}k\left(\mathrm{2+}\lambda \right)\mathrm{-}\mathrm{4}{\mathrm{\ }M}^{\mathrm{2}}_p\mathrm{\ }N\mathrm{\ (1+2}\lambda \mathrm{)}}}{\sqrt{\mathrm{2+}\lambda }}\right)}^k}{\left(\mathrm{1+2}\lambda \right)}}}
\end{equation}

Now, we specify the constraints on the parameters using the values obtained from the observations. Using of the allowed values  that achieved  for the pair of parameters (n, k), at our previous work, \cite{29} , we can plot the r$-$ $n_{s}$
  diagram  as follows
The best value for the parameters are acquired while the maximum fit is obtained on the observation diagram.

Here,  in Fig 1,  the  plot shows a good match between Planck 2013 observations for  free parameters guesstimated

\begin{figure}[H]
\centering
\includegraphics[scale=0.39]{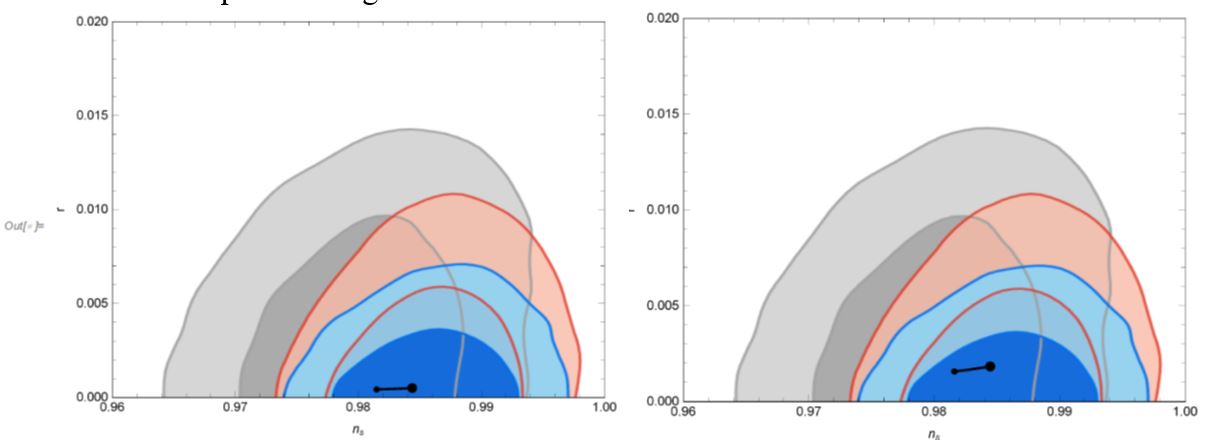}
\caption{
(Color online) The $r-n_s\ $diagram shows prediction of the model in the weak regime for free parametes. In the left panel  $\alpha =70$ , $n=3,\ k=4\ ,\ V_0=2.7\times {10}^{-12},\ \ a={10}^7,\ \ \lambda =1$ , $M_p=1$,  and $N_e=65.\ $In the right panel$\ \alpha =70$ , $n=3,\ \ k=4\ ,\ V_0=2.7\times {10}^{-13},\ \ a={10}^7,\ \ \lambda =5$ , $M_p=1$,  and $N_e=65$  in comparison to the observational data by Planck  2013 , 2015. The likelihood of Planck 2013 is indicated with gray contours, Planck 2015 TT+lowP with red contours and Planck 2015 TT,TE, EE+ lowP with blue contours. in both figures, the thick black lines refer to the predictions of theatrical results in which small and large circles are the values of $n_s$ at the numbers of e-folds N=55, N=65, respectively.
} 
\end{figure}

In Recent observational data, the amplitude of scalar perturbations across the horizon is as ${\mathcal{P}}_{\mathrm{s}}=2.17\pm 0.1\times {10}^{-9}$,and the tensor to scalar ratios $r<0.11$ at 68\%  . ( \cite{43}$-$\cite{47})
has been shown.
 Using the Eqs .($\eqref{33}$ and $\eqref{34}$) \underbar{, we} plot the diagram of the running parameter$\ \frac{dn_{s\ }}{dN}-n_{s\ }$ and compare it with observations. Figure 2, the plot  indicates the prediction of this model, it is clear that lie inside the joint 68\% Cl region of Planck 2015 TT, TE, EE+lowP  data,( \cite{47}$-$\cite{48})
and so could satisfy the compatibility with observations.

\begin{figure}[H]
\centering
\includegraphics[scale=0.6]{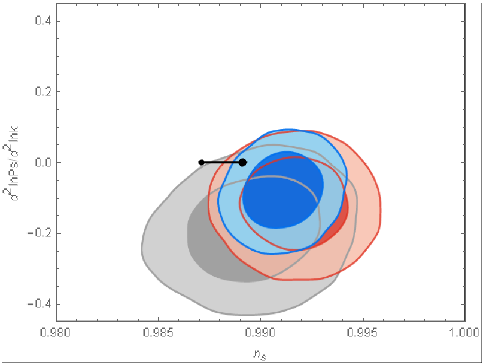}
\caption{(Color online) In this figure, the thick black line depicts the predictions of our model in which small and large circles are the values of $n_s$ at the numbers of e-folds N=55 , N=65, respectively. To plot this shape we use the free parameters $n=3,\ \ k=4\ ,\ V_0=2.7\times {10}^{-13},\ \ \lambda =5$ , $M_p=1$ , ${\mathrm{N}}_{\mathrm{e}}\mathrm{=65}$ and $\alpha =70.\ $In this figure the gray contours indicates the likelihood of Planck 2013, Planck 2015 TT+lowP showed with red contours and the blue contours considered for Planck 2015 TT, TE, EE +lowP.} 
\end{figure}
We know in warm inflation model, the thermal fluctuations overcome the quantum fluctuations. this is an important feature of the warm inflationary model since the fluid temperature is bigger than the Hubble parameter, i.e. T$ >$ H. To receive a healthy warm inflation, this condition should be justified during the cosmological evolution. Figure 3 expresses the behavior of the ratio of the temperature to the Hubble parameter during such era.

\begin{figure}[H]
\centering
\includegraphics[scale=0.6]{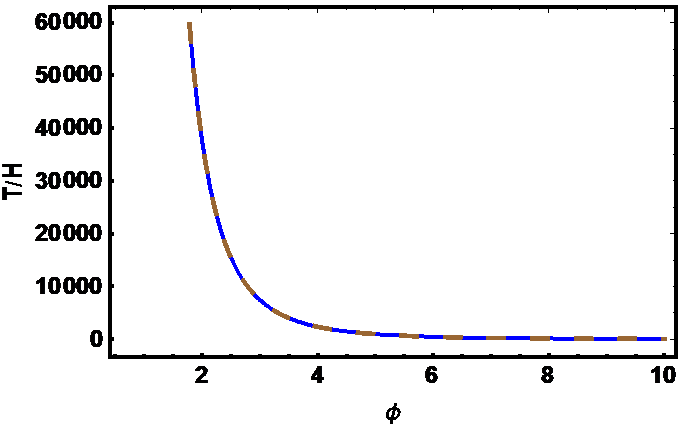}
\caption{(Color online) the plot shows the ratio of the temperature to the Hubble parameter during the inflationary period of the model in the weak dissipative regime. The inflation scalar field,$\phi $ for different values of (n, k) and the parameter a. As one can see from the plots during inflation, the temperature is larger than the Hubble parameter, and the condition T $\mathrm{>}$ H is satisfied properly. To draw these figures, we fixed the value of V0 at $V_0=2.7\times {10}^{-13}$ (blue line:n$=3$;k$=4$;a$=10^{7}$;$\lambda=1$ and red dashed lines:n$=3$;k$=4$;a$=10^{7}$;$\lambda=5$.)} 

\end{figure}

In the following, we examine our model in a strong dissipative regime with the same amount of potential.
\section{	Investigation of Effects Anisotropy in strong Dissipative Regime }
Considering  $V\left(\mathit{\Phi}\right)=V_0{\mathit{\Phi}}^k$ the same as the previous section, we extend our calculations for strong dissipation regimes, for $R\gg 1$

\noindent Using $H^2=\frac{1}{\left(2\lambda +1\right)M^2_P}\ V\left(\mathit{\Phi}\right)$ and $H\ \left(\lambda +2\right)\ R\ \dot{\phi }+V_{,\phi }=0,\ \ $ $R\equiv \frac{\mathrm{\Gamma }}{\left(\lambda +2\right)H_2}$ the temperature of the radiation is gained as

\begin{equation} \label{9} 
T\mathrm{=}\overline{\gamma }\mathrm{\ \ }{\phi }^{\frac{\mathrm{-}\mathrm{3-}\frac{\mathrm{3}k}{\mathrm{2}}\mathrm{+}n}{\mathrm{4+}n}\mathrm{\ \ \ \ \ \ \ \ \ \ \ \ \ \ \ \ \ \ }}\mathrm{\ \ ,\ \ \ \ \ \ \ }\overline{\gamma }\mathrm{=}{\left(\frac{\mathrm{3}}{\mathrm{4}}\right)}^{\frac{\mathrm{1}}{\mathrm{4+}n}}{M^{\frac{\mathrm{1}}{\mathrm{4+}n}}_P\left(\frac{{\sqrt{\left(\mathrm{2}\lambda \mathrm{+1}\right)}k}^{\mathrm{2}}{V_0}^{\frac{\mathrm{3}}{\mathrm{2}}}}{a\mathrm{\ }\alpha \mathrm{\ }\left(\lambda \mathrm{+2}\right)}\right)}^{\frac{\mathrm{1}}{\mathrm{4+}n}} 
\end{equation} 
The slow-roll parameter is   $\epsilon =\frac{k^2M^2_p}{2{\phi }^2}$
\\
Hence
\begin{equation} \label{8} 
R=\ \frac{aM_P{\phi }^{1-n}{\left(\overline{\gamma }\ {\phi }^{\frac{-3-\frac{3k}{2}+n}{4+n}}\right)}^n\ }{\left(\lambda +2\right)\sqrt{\frac{V_0{\phi }^k}{\left(2\lambda +1\right)}}}\ \ \ \ \ \ \ \ 
\end{equation} 
At the end of inflation, we know in the strong regime,  $\epsilon \simeq R$ . In this case, the scalar field :
\begin{equation} \label{7} 
{\phi }_{end}\mathrm{=}{\left(\frac{\left(\lambda \mathrm{+2}\right)\sqrt{\frac{V_0}{\left(\mathrm{2}\lambda \mathrm{+1}\right)}}\mathrm{\ }k^{\mathrm{2}}M_P}{\mathrm{2}{\overline{\gamma }}^n\mathrm{\ }a}\right)}^{\mathrm{-}\left(\left(\frac{n\mathrm{(-3-}\frac{\mathrm{3}k}{\mathrm{2}}\mathrm{+}n\mathrm{)}}{\mathrm{4+}n}\mathrm{+(1-}n\mathrm{-}\frac{k}{\mathrm{2}}\mathrm{)}\right)\mathrm{+2}\right)}, 
\end{equation} 
And also the field  based on the  number of  e-fold  get: 
\begin{equation} \label{6} 
\phi =\ \ \ \sqrt{2kM^2_P\ N+2^{6-k-2n+\frac{2n\ (-3-\frac{3k}{2}+n)}{4+n}}\ {\left(\frac{k^2M_P{\overline{\ \gamma }}^{\ -n}\ (2+\lambda )\sqrt{\frac{V_0}{(1+2\lambda )}}\ }{a}\right)}^{-6+k+2n-\frac{2n(-3-\frac{3k}{2}+n)}{4+n}}}
\end{equation} 
 from \textit{  }${\mathcal{P}}_s=\frac{25\ H^2}{4{\dot{\ \mathit{\Phi}}}^2}{\delta \mathit{\Phi}}^2\ in\ \   \cite{49},\ \mathrm{we\ know\ from\ } $\eqref{6}$)  and\ {\delta \mathit{\Phi}}^2\simeq \frac{K_FT}{2{\pi }^2}$ , $K_F=\sqrt{\mathrm{\Gamma }(\frac{\lambda +2}{3}H}\boldsymbol{)}$, the scalar power spectrum is obtained as

\begin{tiny}
\begin{align} \label{41} 
\nonumber
&{\mathcal{P}}_{\mathrm{s}}\mathrm{=}\mathrm{\{}\mathrm{\ }\frac{\mathrm{25\ }a^{\mathrm{2}}\mathrm{\ \ }\overline{\gamma }\mathrm{\ }{\left(\mathrm{2+}\lambda \right)}^{\mathrm{2}}\mathrm{\ }}{\mathrm{72}\sqrt{\mathrm{3}}\mathrm{\ }k^{\mathrm{2}}M^{\mathrm{2}}_{P\mathrm{\ }}{\pi }^{\mathrm{2}}\left(V_0\mathrm{+2}V_0\lambda \right)}{\left(\mathrm{2}kM^{\mathrm{2}}_{P\mathrm{\ }}N\mathrm{+}{\mathrm{2}}^{\mathrm{-}y}{\left(\frac{j\mathrm{\ }{\overline{\gamma }}^{\mathrm{-}n}}{a}\right)}^{\mathrm{-}y}\right)}^{\frac{\mathrm{1}}{\mathrm{2}}\left(\mathrm{4-}k\mathrm{-}\mathrm{2}n\mathrm{+}\frac{\mathrm{-}\mathrm{3-}\frac{\mathrm{3}k}{\mathrm{2}}\mathrm{+}n}{\mathrm{4+}n}\right)}{\left(\overline{\gamma }{\left(\mathrm{2}k\mathrm{\ }M^{\mathrm{2}}_{P\mathrm{\ }}N\mathrm{+}{\mathrm{2}}^{\mathrm{-}y\mathrm{\ }}{\left(\frac{{\overline{\gamma }}^{\mathrm{-}n}j}{a}\right)}^y\right)}^f\right)}^{\mathrm{2}n}\mathrm{\times } 
\\  
&\sqrt{\left(a(2+\lambda \right)\ {\left(2kM^2_{P\ }N+2^{-y}\left({\left(\frac{{\overline{\gamma }}^{-n}j}{a}\right)}^{-y}\right)\right)}^{\frac{1-n}{2}}\times \frac{\mathrm{1}}{M_P}\sqrt{\frac{V_0{\left(2kM^2_{P\ }N+2^{-y}{\left(\frac{{\overline{\gamma }}^{-n}\ j}{a}\right)}^y\right)}^{\frac{k}{2}}}{\left(1+2\lambda \right)}}\ \ \ \times {\left(\overline{\gamma }\ {(2kM^2_{P\ }N+{\ \ 2}^{-y}\ \ {\left(\frac{{\overline{\gamma }}^{-n}j}{a}\right)}^y)}^f\right)}^n\ }
\end{align} 
\end{tiny}

Where  $y=\frac{4\left(-6+k+2n+kn\right)}{4+n}$ ,  $j=k^2M_P\left(2+\lambda \right)\sqrt{\frac{V_0}{\left(1+2\lambda \right)}}$  and  $f=\frac{-6-3k+2n}{4(4+n)}$
\[\beta \mathrm{=}{\left(\frac{3}{4}\right)}^{\frac{1}{4+n}}{\left(\frac{M_Pk^2{V_0}^{\frac{3}{2}}}{\ \alpha \ \left(\lambda +2\right)\sqrt{\frac{1}{\left(2\lambda +1\right)}}}\right)}^{\frac{1}{4+n}}, x=\left(4-k-2n+\frac{-3-\frac{3k}{2}+n}{4+n}\right),\] 
The scalar spectral index get 
\[n_s\mathrm{(}n,k,N\mathrm{)=} \mathrm{1+}\] 
\begin{equation}\label{4} 
\frac{\mathrm{3\ \times }{\mathrm{2}}^{\frac{\mathrm{-}\mathrm{8+7}n\mathrm{+4}k\left(\mathrm{1+}n\right)}{\mathrm{4+}n}}\mathrm{\ }k\mathrm{\ }M^{\mathrm{2}}_{P\mathrm{\ }}\left(\mathrm{-}\mathrm{10+8}n\mathrm{+3}k\mathrm{(1+}n\right)){\left(\frac{k^{\mathrm{2}}M_P{\overline{\mathrm{\ }\gamma }}^{\mathrm{\ -}n}\mathrm{\ }\left(\mathrm{2+}\lambda \right)\sqrt{\frac{V_0}{\left(\mathrm{1+2}\lambda \right)}}\mathrm{\ }}{a}\right)}^{\frac{\mathrm{24}}{\mathrm{4+}n}}\mathrm{\ }}{\left(\mathrm{4+}n\right)\left({\mathrm{2}}^{\frac{\mathrm{9}n\mathrm{+4}k\left(\mathrm{1+}n\right)}{\mathrm{4+}n}}\mathrm{\ }k\mathrm{\ }M^{\mathrm{2}}_{P\mathrm{\ }}N{\left(\frac{k^{\mathrm{2}}M_P{\overline{\mathrm{\ }\gamma }}^{\mathrm{\ -}n}\mathrm{\ }\left(\mathrm{2+}\lambda \right)\sqrt{\frac{V_0}{\left(\mathrm{1+2}\lambda \right)}}\mathrm{\ }}{a}\right)}^{\frac{\mathrm{24}}{\mathrm{4+}n}}\mathrm{+}{\mathrm{2\ }}^{\frac{\mathrm{20}}{\mathrm{4+}n}}\mathrm{\ \ }{\left(\frac{k^{\mathrm{2}}M_P{\overline{\mathrm{\ }\gamma }}^{\mathrm{\ -}n}\mathrm{\ }\left(\mathrm{2+}\lambda \right)\sqrt{\frac{V_0}{\left(\mathrm{1+2}\lambda \right)}}\mathrm{\ }}{a}\right)}^{\frac{\mathrm{4}\left(k\mathrm{+2}n\mathrm{+}kn\right)}{\mathrm{4+}n}}\right)} ,      
\end{equation} 
And running get \\

$\alpha (n,k,N)= $
\begin{equation}\label{3} 
 \frac{3\ \times 2^{\frac{-8(-1+k+2n+kn)}{4+n}}\ k^2\ M^2_{P\ }\left(-10+8n+3k(1+n)\right){\left(\frac{k^2M_P{\overline{\ \gamma }}^{\ -n}\ \left(2+\lambda \right)\sqrt{\frac{V_0}{\left(1+2\lambda \right)}}\ }{a}\right)}^{\frac{48}{4+n}}\ }{\left(4+n\right){\left(2^{\frac{9n+4k\left(1+n\right)}{4+n}}\ k\ M^2_{P\ }N{\left(\frac{k^2M_P{\overline{\ \gamma }}^{\ -n}\ \left(2+\lambda \right)\sqrt{\frac{V_0}{\left(1+2\lambda \right)}}\ }{a}\right)}^{\frac{24}{4+n}}+{2\ }^{\frac{20}{4+n}}\ \ {\left(\frac{k^2M_P{\overline{\ \gamma }}^{\ -n}\ \left(2+\lambda \right)\sqrt{\frac{V_0}{\left(1+2\lambda \right)}}\ }{a}\right)}^{\frac{4\left(k+2n+kn\right)}{4+n}}\right)}^2},\ \ 
\end{equation} 
The tensor power spectrum is 
\begin{equation}\label{2} 
{\mathcal{P}}_t\left(n,k,N\right)\mathrm{=}\frac{\mathrm{2\ }V_0{\left(\mathrm{2+}\lambda \right)}^{\mathrm{2}}{\mathrm{(2\ }k\mathrm{\ }M^{\mathrm{2}}_{P\mathrm{\ }}N\mathrm{\ }{\mathrm{+\ 2}}^{\frac{\mathrm{-}\mathrm{4}\left(\mathrm{-}\mathrm{6+}k\mathrm{+2}n\mathrm{+}kn\right)}{\mathrm{4+}n}}\mathrm{\ }{\mathrm{(}\frac{k^{\mathrm{2}}M_P{\ \left(2+\lambda \right)\overline{\mathrm{\ }\gamma }}^{\mathrm{\ -}n}\mathrm{\ }\sqrt{\frac{V_0}{\left(\mathrm{1+2}\lambda \right)}}\mathrm{\ }}{a}\mathrm{)}}^{\frac{\mathrm{-}\mathrm{4}\left(\mathrm{-}\mathrm{6+}k\mathrm{+2}n\mathrm{+}kn\right)}{\mathrm{4+}n}}\mathrm{)}}^{\frac{k}{\mathrm{2}}}\mathrm{\ }}{\mathrm{27\ }M^{\mathrm{4}}_{P\mathrm{\ }}{\pi }^{\mathrm{2}}}
\end{equation}  
The tensor to scalar ratio is acquired  as 
\begin{equation} \label{1} 
r\left(n,k,N\right)\mathrm{=}\frac{\mathrm{(16\ }k^{\mathrm{2}}V_0\left(V_0\mathrm{+2}V_0\lambda \right)d^G{\left(\gamma \mathrm{\ }{\left(d\right)}^h\right)}^{\mathrm{-}\mathrm{2}n}\mathrm{)}}{\left(\mathrm{25\ }\sqrt{\mathrm{3}}\mathrm{\ }a^{\mathrm{2}}M^{\frac{3}{2}}_{P\mathrm{\ }}\gamma \sqrt{\mathrm{(}a\mathrm{\ (2+}\lambda \mathrm{)\ }{\left(\gamma d^h\right)}^n{\mathrm{\ }d}^{\frac{\mathrm{1-}n}{\mathrm{2}}}\sqrt{\frac{V_0\mathrm{\ }d^{\frac{k}{\mathrm{2}}}}{\left(\mathrm{1+2}\lambda \right)}}}\mathrm{\ \ }\right)}  ,                   
\end{equation} 
Where
\[\ \ w\mathrm{=}\frac{\mathrm{4}\left(\mathrm{-}\mathrm{6+}k\mathrm{+2}n\mathrm{+}kn\right)}{\mathrm{4+}n}\ \ \ \ ,\ i\mathrm{=}\frac{k^{\mathrm{2}}\left(\mathrm{2+}\lambda \right)M_P{\overline{\mathrm{\ }\gamma }}^{\mathrm{\ }--n}\sqrt{\frac{V_0}{\left(\mathrm{1+2}\lambda \right)}}}{a}\ \ \ ,\ \ d\mathrm{=}\left(\mathrm{2\ }k\mathrm{\ }M^{\mathrm{2}}_{P\mathrm{\ }}N\mathrm{+}{\mathrm{2}}^{\mathrm{-}w}i^w\right)\ \ ,\ \ \] 
\[\ \ G\mathrm{=}\frac{\mathrm{-}\mathrm{26+6}n\mathrm{+4}n^{\mathrm{2}}\mathrm{+}k\mathrm{(19+4}n\mathrm{)}}{\mathrm{4(4+}n\mathrm{)}},   h\mathrm{=}\frac{\mathrm{-}\mathrm{6-3}k\mathrm{+2}n}{\mathrm{4(4+}n\mathrm{)}}\] 
Now we evaluate   our model based on the results of the observations and apply some constraint to  gain  the parameters that create the best fit.\\

{Using of the allowed values $\boldsymbol{\mathrm{{}}}$$\boldsymbol{\mathrm{{}}}$ that achieved  for the pair of parameters (n, k), at our previous work, }${\boldsymbol{\mathrm{\ }} ref.\cite{29}  \boldsymbol{]}}${ , we can re-plot the }$\boldsymbol{r}\boldsymbol{-}{\boldsymbol{n}}_{\boldsymbol{s}}\boldsymbol{\ }$\textbf{diagram  as follows}

\begin{figure}
\centering
\includegraphics[scale=0.6]{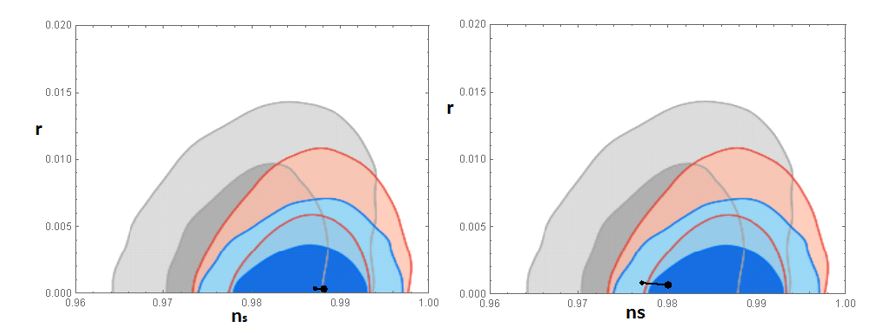}
\caption{(Color online) this is $\frac{dn_s}{dN}-n_s$diagram which explains the running of the parameter $n_s.$ In this figure, the thick black line despicts the predictions of our model in which small and large circles are the values of $n_s$ at the numbers of e-folds$\boldsymbol{\ }N=55$ and $N=65$, respectively. To plot this shape we use the free parameters $n=\frac{1}{9},\ \ k=\frac{1}{7}\ ,\ V_0=2.98\times {10}^{-13},\ \ \lambda =4\times {10}^{-6}$ , $M_p=1$ , $N_e=65$ and $\alpha =70.\ $In this figure the gray contours indicates the likelihood of Planck 2013, Planck 2015 TT+lowP showed with red contours and the blue conturs considered for Planck 2015 TT, TE, EE +lowP.} 
\end{figure}

The best parameters for this diagram are those create the desired match. In Figure 5, we guesstimated some constraint over the parameters of our model compared to the observations.
\\
Therefore, in both regimes, for the potentials of the power-law in the context of anisotropic warm inflation, the suitable agreement with the observations is sighted.
\begin{figure}[H]
\centering
\includegraphics[scale=0.6]{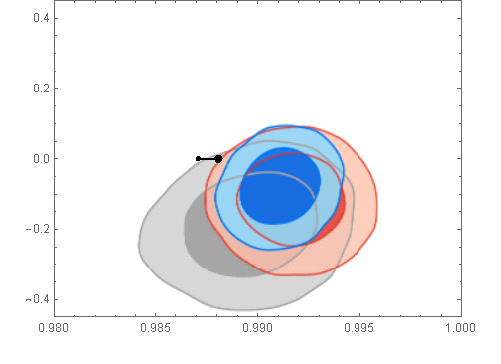}
\caption{
(Color online) this is $\frac{dn_s}{dN}-n_s$diagram which explains the running of the parameter $n_s.$ In this figure, the thick black line despicts the predictions of our model in which small and large circles are the values of $n_s$ at the numbers of e-folds$\boldsymbol{\ }N=55$ and $N=65$, respectively. To plot this shape we use the free parameters $n=\frac{1}{9},\ \ k=\frac{1}{7}\ ,\ V_0=2.98\times {10}^{-13},\ \ \lambda =4\times {10}^{-6}$ , $M_p=1$ , $N_e=65$ and $\alpha =70.\ $In this figure the gray contours indicates the likelihood of Planck 2013, Planck 2015 TT+lowP showed with red contours and the blue conturs considered for Planck 2015 TT, TE, EE +lowP.
}
\end{figure}

The same as the previous section,  the diagram  T/ H versus  is plotted, i.e. T $>$ H.
\begin{figure}[H]
\centering
\includegraphics[scale=0.6]{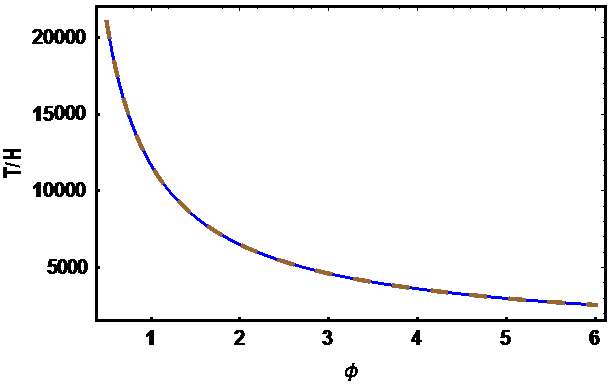}
\caption{
(Color online) the plot shows the ratio of the temperature to the Hubble parameter during the inflationary period of the model in the strong dissipative regime. The inflation scalar field,$\boldsymbol{\phi }$\textbf{ for different values of (n, k) and the parameter a. As one can see from the plots during inflation, the temperature is larger than the Hubble parameter, and the condition T ${\mathrm{>}}$ H is satisfied properly. To draw these figures, we fixed the value of V0 at ${{V}}_{{0}}{=}{2}.{98}{\times }{{10}}^{{-}{13}}$
}(blue line:n$=1.9$;k$=1.7$;a$=10^{-8}$;$\lambda{=}{4}{\times }{{10}}^{{-}{6}}$ and red dashed lines:n$=1.9$;k$=1.7$;a$=10^{-8}$;$\lambda{=}{4}{\times }{{10}}^{{-}{5}}$).
}
\end{figure}

\section{Conclusions}
{Following the research on our article entitled " Constraints on warm power-law inflation in light of Planck results"[29] We saw  which for (n = 2, k = 6) with  V0 = 2.7 $\boldsymbol{\mathrm{\times}}$}${\boldsymbol{\mathrm{10}}}^{\boldsymbol{\mathrm{-}}\boldsymbol{\mathrm{25}}}$\textbf{ in the weak dissipation regime, we have the most agreement with Planck's observational data in 2013 and 2015. In the strong dissipation regime in the isotropic warm inflation model for (n = 1, k = 4)  with V0 = 5 $\boldsymbol{\mathrm{\times}}$}${\boldsymbol{\mathrm{10}}}^{\boldsymbol{\mathrm{-}}\boldsymbol{\mathrm{20}}}$\textbf{  We obtained the most agreement with Planck's observations.. In this work, we have studied the effects of anisotropy on warm inflation in the framework of the Bianchi type I metric and examine the model for the power-law potential in two models of the weak and strong dissipative regime, and the different $\boldsymbol{\mathrm{{}}}$$\boldsymbol{\mathrm{{}}}$ parameters of slow- roll and power spectrum of scalar and tensor perturbations, scalar and tensor spectral indices, their running and tensor-to-scalar ratio were been gained. Using of the allowed values of the pair }$\boldsymbol{(}\boldsymbol{n},\boldsymbol{k}\boldsymbol{)}$\textbf{ in the reference [29],  and Plotting  the }$\boldsymbol{r}\boldsymbol{-}{\boldsymbol{n}}_{\boldsymbol{s}}\boldsymbol{\ }$\textbf{diagram and its running, it  has been estimated the best regions compatible  with the observations for the free parameters.In the weak dissipation regime the pair (}$\boldsymbol{n}\boldsymbol{=}\boldsymbol{3}\boldsymbol{,\ }\boldsymbol{k}\boldsymbol{=}\boldsymbol{4}$\textbf{) for }$\boldsymbol{r}\boldsymbol{-}{\boldsymbol{n}}_{\boldsymbol{s}}$\textbf{ diagram with }$\boldsymbol{\lambda }\boldsymbol{=}\boldsymbol{1},\boldsymbol{5}$\textbf{ .it has been seen the best fit with the observations  that  shown it in Figures 1 and 2 . These figures have  illustrated  that our results are consistent with Planck's observations. The same argumentation}  is true for the strong dissipative regime in the presence of the parameters ($n=\frac{1}{9},\ k=\frac{1}{7}\ $) with, $V_0=2.98\times {10}^{-13}\ \ $and $\ \lambda =4\times {10}^{-6}$ . For this regime, due to presence of the parameter  . the chance in the selection of free parameters $(n,k)$ is less than the weak case. We will continue this study in the future by examining viscosity.
\\
{The new consequences of the anisotropic scenario against the isotropic setup is shown in Fig.1, that is left  panel related to }$\boldsymbol{\lambda }\boldsymbol{=}\boldsymbol{1},${  to be   the isotropic and right panel }$\boldsymbol{\lambda }\boldsymbol{=}\boldsymbol{5}\boldsymbol{,\ }${ is anisotropic setup }. 

\section*{
Acknowledgments}
Zahra Ghadiri would like to thanks M. Naderi for her guidance. And she is also thanks her sister N. Nili for the good helps and she specially thanks go to her mother M.dabbaghha for her permanent support

\end{document}